# Machine Learning Assisted Design and Optimization of Transition Metal-Incorporated Carbon Quantum Dot Catalysts for Hydrogen Evolution Reaction

Duong Nguyen Nguyen,[1] Min-Cheol Kim,[1] Unbeom Baeck,[1] Jaehyoung Lim,[2] Namsoo Shin,[3] Jaekook Kim,[4] Heechae Choi,[5] Ho Seok Park,[1,6]*, Uk Sim,[2,7,]* and Jung Kyu Kim[1,]*

[1]School of Chemical Engineering, Sungkyunkwan University (SKKU), 2066, Seoburo, Jangan-gu, Suwon 16419, Republic of Korea
[2]Hydrogen Energy Technology Laboratory, Korea Institute of Energy Technology (KENTECH), Naju 58330, Republic of Korea
[3]Deep Solution Inc., 17, Jukjeon ro, Giheunggu, Yonginsi, Gyeonggido 16897, Republic of Korea
[4]Department of Materials Science and Engineering, Chonnam National University, Gwangju, 61186, Republic of Korea
[5]Theoretical Materials & Chemistry Group, Institute of Inorganic Chemistry, University of Cologne, Greinstr. 6, 50939 Cologne, Germany
[6]SKKU Institute of Energy Science and Technology (SIEST), Sungkyunkwan University (SKKU), Suwon, 16419, Republic of Korea
[7]Research Institute, NEEL Sciences, INC., Gwangju, 61186, Republic of Korea

**Correspondence**
Jung Kyu Kim, School of Chemical Engineering, Sungkyunkwan University (SKKU), 2066, Seoburo, Jangan-gu, Suwon 16419, Republic of Korea
Email: legkim@skku.edu

Uk Sim, Hydrogen Energy Technology Laboratory, Korea Institute of Energy Technology (KENTECH), Naju, Republic of Korea. Research Institute, NEEL Sciences, Inc., Naju, Republic of Korea
Email: usim@kentech.ac.kr

Ho Seok Park, School of Chemical Engineering, SKKU Institute of Energy Science and Technology (SIEST), Sungkyunkwan University (SKKU), Suwon, 16419, Republic of Korea
Email: phs0727@skku.edu

**Funding information**
This work was supported by the National Research Foundation of Korea (NRF) grant funded by the Korean government (MSIT) (NRF-2022R1A2C1011559).

||Duong Nguyen Nguyen and Min-Cheol Kim contributed equally to this work.

Development of cost-effective hydrogen evolution reaction (HER) catalysts with outstanding catalytic activity, replacing cost-prohibitive noble metal-based catalysts, is critical for practical green hydrogen production. A popular strategy for promoting the catalytic performance of noble metal-free catalysts is to incorporate earth-abundant transition metal (TM) atoms into nanocarbon platforms such as carbon quantum dots (CQDs). Although data-driven catalyst design methods can significantly accelerate the rational design of TM element-doped CQD (M@CQD) catalysts, they suffer from either a simplified theoretical model or the prohibitive cost and complexity of experimental data generation. In this study, we propose an effective and facile HER catalyst design strategy based on machine learning (ML) and ML model verification using electrochemical methods accompanied with density functional theory (DFT) simulations. Based on a Bayesian genetic algorithm (BGA) ML model, the Ni@CQD catalyst on a three-dimensional reduced graphene oxide (3D rGO) conductor is proposed as the best HER catalyst under the optimal conditions of catalyst loading, electrode type, and temperature and pH of electrolyte. We validate the ML results with electrochemical experiments, where the Ni@CQD catalyst exhibited superior HER activity, requiring an overpotential of 189 mV to achieve 10 mA cm$^{-2}$ with a Tafel slope of 52 mV dec$^{-1}$ and impressive durability in acidic media. We expect that this methodology and the excellent performance of the Ni@CQD catalyst provide an effective route for the rational design of highly active electrocatalysts for commercial applications.

# 1. INTRODUCTION

Hydrogen is a renewable, green, and sustainable energy carrier and has become a potential ideal clean energy source.[1-3] Among various hydrogen generation methods, electrochemical water splitting is one of the most promising approaches for producing clean $H_2$ with high purity.[4-5] However, water splitting under alkaline conditions suffers from the high energy barrier of the O–H bond cleavage reaction, whereas, under acidic conditions maintaining the stability of the electrocatalysts is difficult, making electrocatalytic $H_2$ production challenging for industrialization.[6-7] To achieve high efficiency in the hydrogen evolution reaction (HER), noble metal-based catalysts such as Pt, Pd, and Ru should be used in acidic conditions.[8-10] However, the most active noble metal-based catalysts cannot be widely used due to their scarcity, high cost, and lack of stability. Therefore, for the commercialization of the HER, it is essential to develop highly efficient and durable non-precious-metal-based catalysts. To realize such efficient and durable earth-abundant transition metal (TM) catalysts, the remained challenge is to promote the catalytic activity and stability of these catalysts.

A common strategy for enhancing the catalytic performance of a metal catalyst is to confine the metal cluster size to a sub-nm scale by incorporating metal catalysts into a support material with a confined dimension.[11-12] As the metal cluster size decreases, the portion of undercoordinated metal sites (or edge sites) increases, where such undercoordinated metal sites are highly active catalytic sites.[13] Therefore, the number of active site per total metal atom number increases as the metal cluster size decreases.[14] Moreover, a decreased metal cluster size can improve the quantum confinement effect on free electrons within the metal cluster,[15-16] which alters the electronic structure and can be beneficial for catalysis in terms of metal–adsorbate binding energy. Recently, carbon quantum dots (CQDs), which refer to carbon nanomaterials with a sub-10 nm size, have been extensively used as support materials that can restrain the size of TM catalysts.[17] CQDs have an abundance of highly active intrinsic vacancies/defects, which can strongly bind with TM atoms or can easily adopt functional groups that can strongly bind with TM atoms.[18-20] Thus, TM catalysts are easily incorporated into the carbon network and avoid aggregation during synthetic and catalytic processes.[21-23] In this way, the association of CQDs with TM catalysts to form an M@CQD (M = transition metals) structure can provide a potential route for engineering non-precious metal-based HER electrocatalysts. Hence, it is critical to discover the optimal TM element that enables an efficient and stable M@CQD HER catalyst.

Recently, with the emergence of high-performance computing devices, high-throughput screening with density functional theory (DFT) simulations and machine learning (ML) models trained on DFT data have been extensively used to predict candidate materials for catalysts.[24-29] Although such data-aided catalyst screening strategies can be a useful tool for accelerating the design of efficient electrocatalysts, realizing the predicted candidate materials as practical electrocatalysts is challenging due to the over-simplification of DFT models and the complexity of electrochemical systems.[30-31] The performance of an electrocatalyst is not only solely determined by its intrinsic properties but is also affected by many experimental factors such as the type of precursor, substrates, pH, temperature, electrodes, electrolytes, catalyst loading, and amount of metal dopants.[32-34] Implementation of such experimental factors to a DFT-based prediction model (or DFT data-driven ML prediction models) is either cost-prohibitive or too convoluted, which further complicates the situation. Thus, for the rational design of practical and efficient electrocatalysts, it is critical to use a prediction model that can optimize various experimental factors, including the type of catalyst (which would be the type of element of the TM dopant in the M@CQD catalyst).[35] The traditional practice to optimize such experimental conditions would be using the one variable at a time (OVAT) method which involves repeating experiments with all variables held constant except one variable of interest being changed until the target property is optimized, and this process is repeated for all variables. The OVAT method is only effective when all variables are independent, which is not the case in electrochemical experiments.[35] To overcome such limitations, more sophisticated methods such as the design of experiments (DOE) are commonly used.[36-37] Unfortunately, DOE methods suffer from poor scaling with increasing dimensions requiring copious amounts of data for convergence. ML techniques are effective for such nonlinear multivariable problems,[38] and recent ML models built with high-throughput experimental data from autonomous lab experiments[39] or continuous flow experiments[40] exibit promising results for targeted nanoparticle synthesis. Such an process is a powerful tool for optimization, yet the complexity and cost of building such high-throughput instruments are not trivial, particularly for state-of-the-art electrochemical experiments.

In this work, we present a simple and facile catalyst prediction and experimental condition optimization strategy that can be readily used in any standard electrochemistry laboratory by combining an ML-based catalyst prediction and optimization step with experimental and theoretical verification steps for ML prediction. We chose Ni, Co, Fe, Mn, Zn, and Ag TM catalysts supported by the nitrogen-doped CQDs as a representative group of M@CQD to investigate their potential HER activity. Due to the complexity of the electrochemical system, ML techniques were first used to predict catalytic properties of these M@CQDs to optimize the experimental conditions and find the catalyst with the best performance as well. Several factors influencing catalytic performance were considered input data, including conductors, loading amount, electrode type, temperature, and pH. We used the genetic algorithm (GA), which is advantageous for a mutivariable optimization problems with a limited amount of prior knowledge,[41] and the Naïve Bayes classifier[42] during the selection process to accelerate the convergence. The trained model accurately predicts optimized electrochemical target values using predicted input variables from unknown space, *e.g.,* experimental conditions not provided from prior knowledge. The prediction is validated by obtaining experimental data under the predicted optimized conditions, which demonstrates that the ML process can predict the results of unexplored experimental conditions with high accuracy and provide a complete picture of M@CQD HER performance within the parameter space. Finally, DFT calculation were used to gain insight into how the TM dopant enhances the HER activity. This methodology demonstrates how a combined strategy of an ML-based data-driven prediction model and experimental and theoretical verification steps can pave the way toward efficient electrocatalytic models.

## 2. RESULTS AND DISCUSSION

### 2.1. Machine learning algorithm to trace and predict M@CQD electrocatalysis

In this work, we use a global search technique, *i.e.*, the GA, to explore the catalytic performance of the M@CQDs which has been demonstrated to be effective in non-linear multivariable optimization problems with minimum prior knowledge.[41] The principle of the GA is to optimize an objective function to solve complex problems by iteratively refining a population of solution candidates using biological inspiration as selection, breeding, and mutation. For the GA to be effective over random features, we consider all input variables to be categorized, *i.e.*, discretized, based on Bayes' theorem to minimize the dimension and significantly reduce the cumulative number of experiments. A flowchart on the working principle of the proposed Bayesian GA (BGA) used in this study is depicted in Figure 1a. The part marked in blue corresponds to the experimental optimization process by the conventional GA, and only the data selected so far among the data selected by crossover/mutation operations have been added to a new batch.

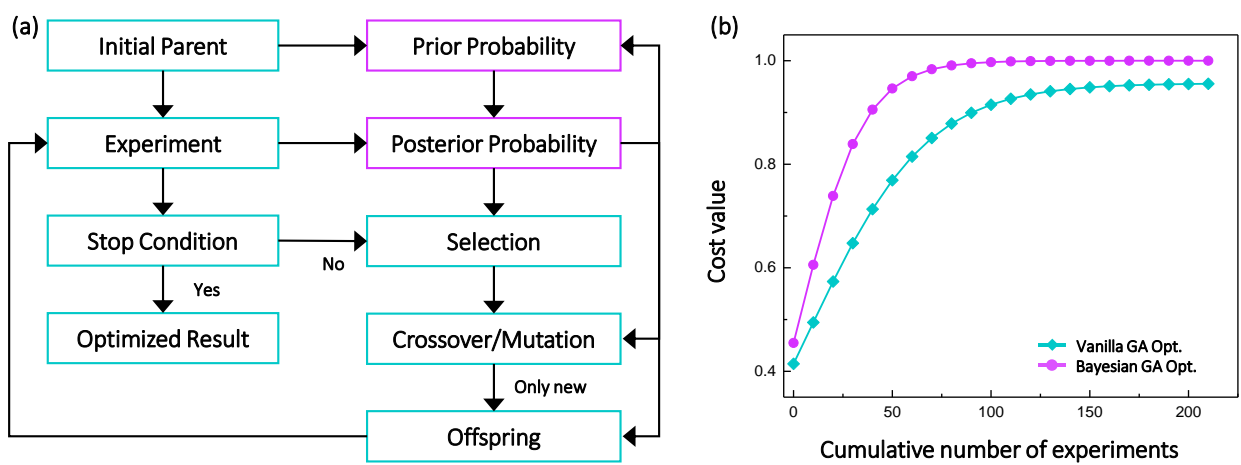

**Figure 1.** (a) Flowchart of the working principle of the proposed BGA. (b) Performance comparison between conventional GA optimization and BGA optimization for a generic convex-shaped cost function.

The efficiency of the BGA optimization is investigated by comparing it with the conventional GA (*i.e.*, "Vanilla" GA) optimization. For accurate performance comparison, it is necessary to compare various cost functions, however, in this work, a simple convex-shaped cost function was used, as such a function is commonly observed in actual experiments. A convex-shaped cost function with a maximum of 1.0 was chosen for six categorical variables with 10 levels each. The GA and BGA are random processes; thus, the averaged results of 2,000 optimization processes are shown in Figure 1b. The simulation results show that the BGA reaches optimal performance in approximately 100 cumulative experiments, whereas the conventional GA does not reach optimal performance even after 200 cumulative experiments. For a simple convex-shaped cost function, which is common for a general electrochemical experiment, the BGA can find optimal conditions with fewer cumulative experiments than the conventional GA.

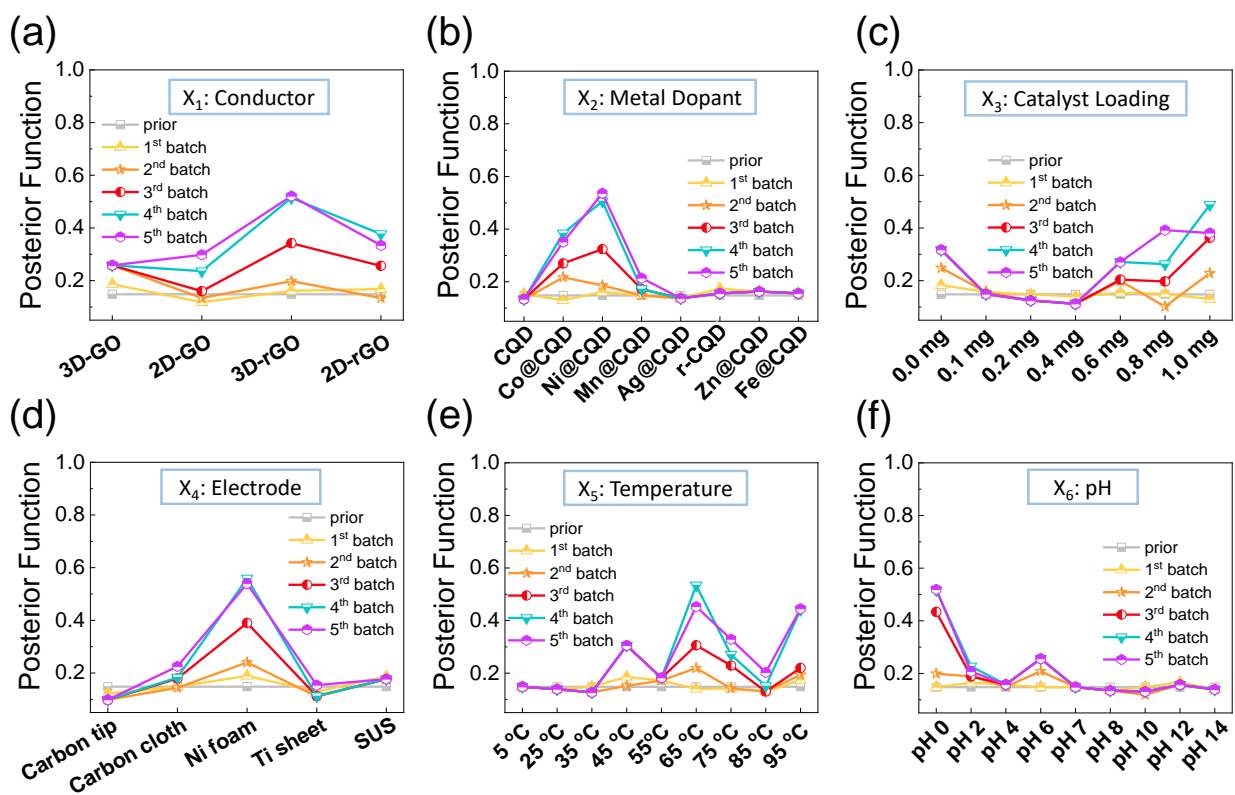

**Figure 2.** Evolution of posterior probability with the experiment batch for the experimental parameters of electrochemical measurement, including (a) conductors, (b) TM dopants, (c) loading amount of electrocatalyst, (d) substrate electrode, (e) operating temperature, and (f) pH of the electrolyte.

The method described above was used to search for the optimal conditions for the water splitting experiment. We search for the catalyst and experimental conditions with optimal overpotential, Tafel slope, and exchange current density. Here, a multi-objective algorithm is used. Figure 2 shows the optimization of experimental condition variables and levels. Here, two dimensional (2D) and three dimensional (3D) graphene oxide (GO) and reduced graphene oxide (rGO) conductors are investigated (Figure 2a); N-doped CQD with oxidized edges (CQD), reduced N-doped CQD with relatively few oxidized edges (r-CQD), and M@CQD electrocatalysts are investigated (Figure 2b).

The cost function is expressed as:

$$Cost = \frac{1}{3}\left\{\frac{(OP+1.6)}{1.6} - \frac{(TS-100)}{80} + \frac{(ECD+5)}{3}\right\}$$

where OP denotes overpotential and has a value of −1.6–0.0 (V), TS denotes a Tafel slope with a value of 20–100 (mV dec$^{-1}$) and ECD represents exchange current density and has a value of −5–−2 (log$_{10}$ A cm$^{-2}$).

The initial prior probability has a uniform distribution with the averaged value of the cost function of the first randomly selected batch. Here, the sum of the probability distributions was not normalized to directly advance the evolution of the posterior probability with the value of the cost function normalized between 0 and 1. In the simulation the convergence speed was the fastest under these conditions.

The first batch contained 12 data sets. From the first batch, we obtained 12 cost function values and selected 2 data sets that maximize the cost function and included them in the next batch. Then, we constructed 12 data sets for the next batch by selecting 5 data sets through a uniform crossover operation and 5 data sets through a mutation operation. Process conditions such as batch size, crossover and mutation ratio were optimized to minimize the cumulative number of experiments through simulation.

The posterior probability was calculated using the following equation.

$$P_{post}(x_i)^{n+1} = cost^n \cdot \alpha + P_{post}(x_i)^n \cdot (1-\alpha).$$

An α value of 0.2 was used, which was confirmed as the fastest convergence condition in the simulation. The convergence of the experimental optimization process can be set by various conditions such as the convergence of the maximum value of the cost function, however, it is most practical to look at the evolution of posterior probability at the BGA optimization. Stopping the evolution of the posterior probability means that no additional information is obtained through experimentation, particularly in the search for an optimal value of the cost function will be stopped. In this work, after examining the evolution of posterior probability, the cumulative number of experiments was minimized by stopping the process when the average absolute change in posterior probability < 0.035.

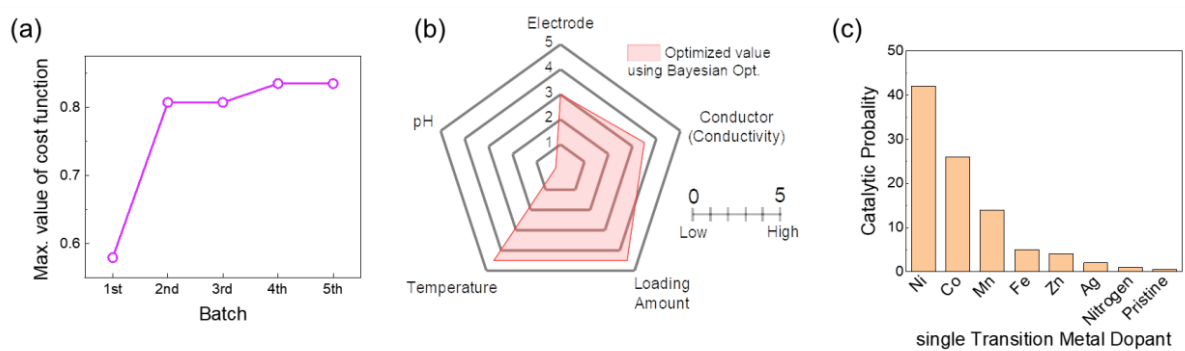

**Figure 3.** Output results of the trained model using BGA optimization in recognizing catalytic performance of M@CQDs under various conditions. (a) Maximum cost function of each batch. (b) Score evaluation of important variables for electrochemical measurement. The value of electrode is assigned to carbon tip (1), carbon cloth (2), nickel foam (3), Ti sheet (4), and stainless-steel SUS (5). (c) Ranking of TM dopant in CQD toward catalytic performance.

The change in posterior probability of each variable and the change in the maximum cost function for each batch were considered. The convergence criteria were met in the fifth batch; thus, the optimization process was stopped (Figure 3a and S1). The experimental conditions that provide the maximum cost are shown in Figure 3b. According to the result, Nickel is the optimum dopant in CQDs toward the HER (Figure 3c) when the catalyst loading is 0.8 mg and combined with 3D rGO as the conductor and Ni foam as the electrode at a temperature of 65°C and pH 0. Hence, we further conducted experiments and atomic theoretical calculations to confirm whether a Ni@CQD would be the best catalyst.

## 2.2. Material characterization

*2.2.1. Morphology and chemical properties*

The TM-doped CQD (hereafter, M@CQD, with M denoting the TM catalyst) was synthesized using fumaronitrile and metal acetate as carbon and metal sources (except Ag@CQD using silver nitrate), respectively, under facile hydrothermal conditions. For comparison, the synthesis of CQDs without using a metal precursor was also performed via a hydrothermal route (CQD) and a solvothermal route using tetrahydrofuran (THF) as a solvent (r-CQD). Herein, we chose fumaronitrile as the oxygen-free nitrogen-containing precursor for the synthesis of CQD which was first reported in 2016 and explained in the formation mechanism in 2021 by the Bae group.[43-44] The unique formula of fumaronitrile can ensure well-ordered edges terminated by $sp^2$-bonded nitrogen atoms and thus can coordinate with metal ions with empty d orbitals to form chemically stable CQD–metal bonds.[45-46] As shown in Figure 4a, the self-assembly growth of CQDs using the hydrothermal route would generate a carbon network with relatively high oxygen content, whereas the solvothermal synthesis process prohibits the formation of oxygen species at the edge of r-CQD. Meanwhile, the complete self-assembly reaction of fumaronitrile in the presence of a small amount of a TM would induce the doping of the TM into the defect of the CQD via two distinctive but simultaneous reactions: the coordination of the TM to the fumaronitrile monomer as an electrophile in the initial step (reaction 1) and the insertion of the TM into the carbon network during the growth process of the CQD (reaction 2).[47] The molar ratio of metal-to-precursor was fixed at 3% to prevent any side reaction. No photoluminescence (PL) was detected in the product when a high amount of metal precursor (*i.e.*, >15%) was used. Such a high amount would result in the reaction yielding undesirable products such as metal hydroxyl rather than quantum dots (Figure S2), whereas a medium amount would induce a mixture of both.

Given the experimental conditions predicted by the ML model, we investigated the morphology of Ni@CQDs. The transmission electron microscopy (TEM) image in Figure 4b and high-angle annular dark field scanning TEM (HAADF-STEM) image in Figure 4c shows that Ni@CQDs are uniformly dispersed without discernible aggregation with an average size of 3.2 nm (see the inset bar-chart of the particle size distribution). This average size of Ni@CQDs is similar to the value of the pristine CQD (Figure S2c), indicating that a small amount of the metallic dopant is incorporated into the CQD and does not affect the morphology. The high-resolution TEM (HRTEM) image in Figure 4d shows a well-defined lattice fringe with a lattice spacing of 0.21 nm, corresponding to the (100) plane of the graphitic domain. As shown in Figure S3, energy-dispersive X-ray spectroscopy (EDX) was performed in Figure S3, indicating that the Ni element is distributed throughout the structure. In addition, the ratio of nickel to nitrogen atoms in Ni@CQD was estimated to be 66%, which is consistent with the homogeneous distribution of nickel in the carbon network via Ni-N bonds. According to the element analysis from X-ray photoelectron spectroscopy (XPS) characterization (Figure 4e), small amounts of the TM are observed in the M@CQD catalysts.

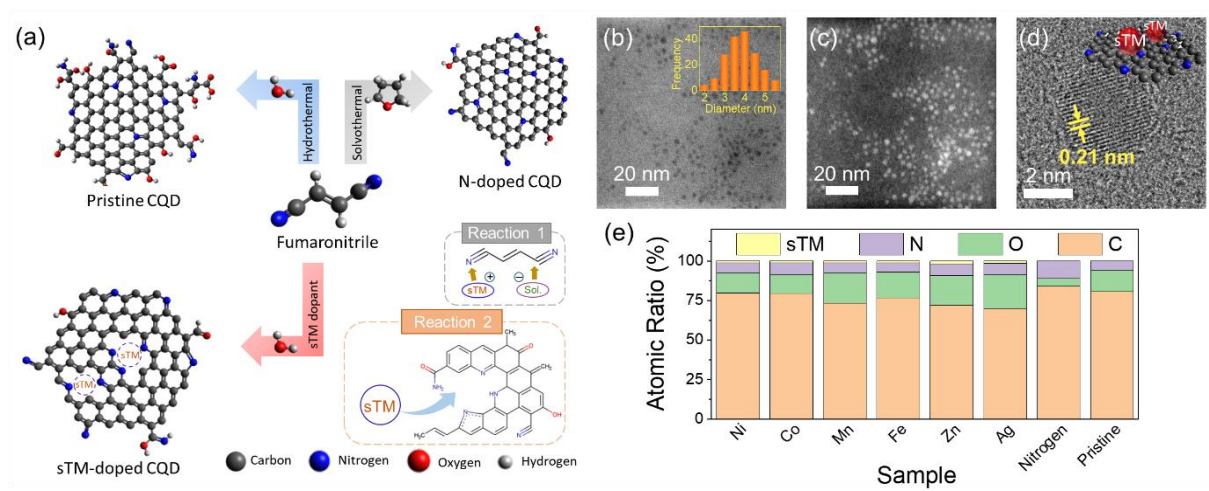

**Figure 4.** Synthesis and morphology of M@CQD. (a) Suggested formation mechanism of M@CQDs through the thermolytic self-assembly reaction of fumaronitrile. (b) Low magnification TEM with a size distribution, (c) HAADF-STEM, and (d) high magnification TEM of Ni@CQD. (e) Elemental analysis from XPS measurements of M@CQDs.

X-ray diffraction (XRD) patterns of the Ni@CQD are shown in Figure S4, showing that the nickel doping does not affect the crystal structure of the CQDs, with a broad reflection at ca. 22° corresponding to the (100) lattice facet of the CQDs. XPS analysis was also performed to analyze the detailed chemical composition of M@CQDs. Figures 5a-c present the Ni 2p, C 1s, and N 1s high-resolution spectra for Ni@CQD. The binding energies of Ni@CQD at 855.1 and 856.4 eV are attributed to Ni $2p_{1/2}$ of $Ni^0$ and $Ni^{2+}$, respectively, whereas the high binding energies at 872.6 and 874.7 eV are attributed to Ni $2p_{3/2}$ of $Ni^0$ and $Ni^{2+}$, respectively. The C 1s and N 1s high-resolution peaks of Ni@CQD and the pristine CQD are shown in Figures 5b-c. Compared with that of the pristine CQD structure, the C 1s binding energy of the Ni@CQD catalyst exhibited significant oxidized carbon peaks. Meanwhile, the N 1s spectrum of Ni@CQD represented a typical enhancement of pyrrolic N and a weak peak of oxidized N. Accordingly, the Ni doping may have affected the CQD oxidation state via simultaneous Ni–C and Ni–N bond formation within the M@CQD network. Similar phenomena in the XPS spectrum were observed in other CQD samples doped with other TMs, as shown in Figures S5–10.

Figure 5d shows the Fourier transform infrared (FT-IR) spectra of the pristine CQDs and CQDs doped with Ni, Co, and Mn. The vibrations at 3423 $cm^{-1}$ (N-H/O-H), 2216 $cm^{-1}$ (-C≡N), 1632 $cm^{-1}$ (H-O-H), 1408 $cm^{-1}$ (C-O/C-H), and 1043 $cm^{-1}$ (C-C) agree well with the XPS results. UV-Vis and PL spectra were studied to account for the optical properties of CQDs. The UV-Vis spectra exhibit absorption peaks at 278 nm with a significant shoulder at 385 nm which are a characteristic feature of C-N/C-O bonds and defect/edge states, respectively. The significant enhancement of these absorption peaks in M@CQD, compared with the pristine CQD, demonstrates the metal doping effect on the electronic transition states *via* metal–carbon and metal–nitrogen bonds. The PL emission spectra of CQDs were obtained within the excitation wavelength range of 340–460 nm (Figure 3f), confirming the characteristic optoelectronic property of quantum confinement of the as-prepared samples. FT-IR and optical spectra of other samples are shown in Figures S11-12.

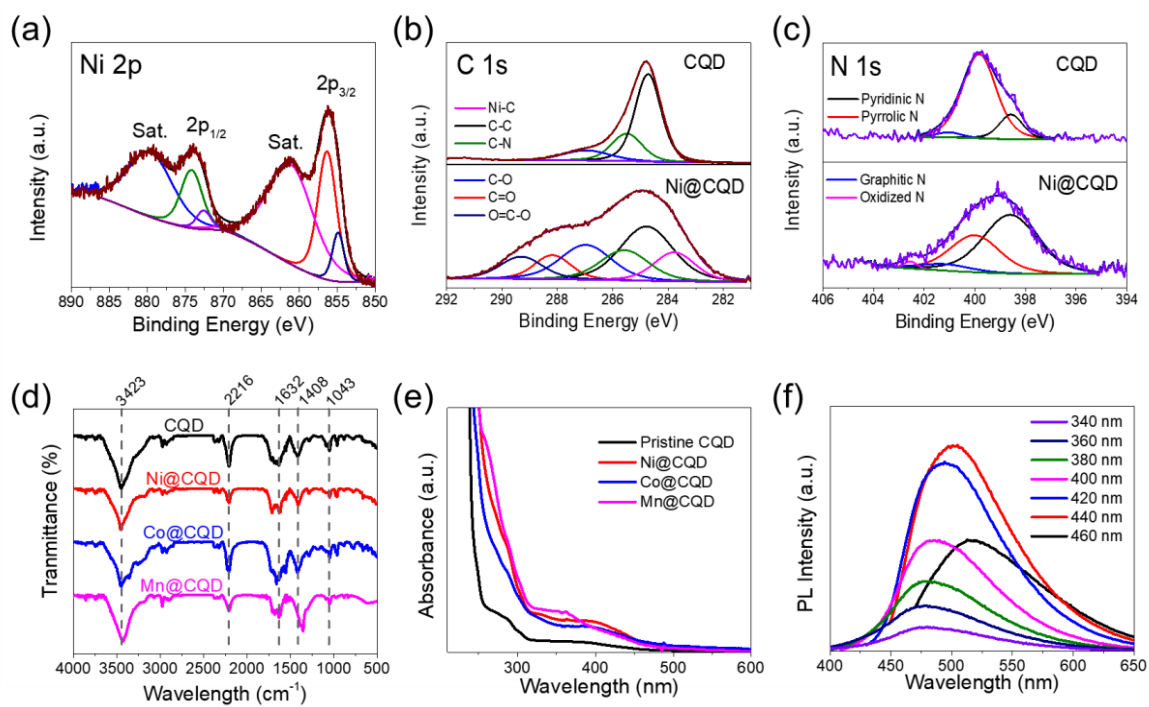

**Figure 5.** Chemical characterization of M@CQD. High-resolution XPS spectra of (a) Ni 2p, (b) C 1s, and (c) N 1s for Ni@CQD. (d) FT-IR, and (e) UV-Vis spectra of pristine CQD, Ni@CQD, Co@CQD, and Mn@CQD. (f) PL spectra of Ni@CQD.

*2.2.2. Electrocatalytic performance*

The as-prepared CQD/3D rGO catalysts were dispersed in a mixture of ethanol and distilled water, removing any agglomerated precipitation, and dropped (~40 μg) onto a nickel foam electrode for electrochemical characterization. Figure 6a shows linear sweep voltammetry (LSV) curves of the three best performing M@CQD/3D rGO electrocatalysts (M = Ni, Co, and Mn), Pt/C, and pristine CQD/3D rGO in 0.5 M $H_2SO_4$ electrolyte with iR compensation. The pristine CQD catalyst exhibits poor HER performance which requires an overpotential ($\eta$) of approximately 390 mV vs. reversible hydrogen electrode (RHE) to reach $j$ of 10 mAcm$^{-2}$. Meanwhile, all M@CQD catalysts exhibit enhanced HER activity than the pristine CQD catalysts. The Ni@CQD catalyst exhibits the lowest overpotential among the M@CQD catalysts of $\eta_1 = 135$ mV and $\eta_{10} = 189$ mV vs. RHE required to generate a current density of 1 and 10 mA cm$^{-2}$, respectively. This result demonstrates that the Ni dopant is the most beneficial for the HER. The corresponding Tafel slope shown in Figure 6b provides a small value of 52 mV dec$^{-1}$ for Ni@CQD, higher than that for Pt/C (36 mV dec$^{-1}$) but significantly lower than those for Co@CQD (64 mV dec$^{-1}$), Mn@CQD (60 mV dec$^{-1}$), and particularly pristine the CQD (78 mV dec$^{-1}$), revealing fast HER kinetics derived from the Ni dopant. Overall, the HER performance of the Ni@CQD sample is the best among the considered M@CQD electrocatalysts (Figure 6c, S13). Furthermore, the overpotential ($\eta_1 = 135$ mV), Tafel slope (52 mV dec$^{-1}$), and exchange current density ($i_0 = 3.7 \times 10^{-2}$ mA cm$^{-2}$) of Ni@CQD is sigficantly improved than those values ($\eta_1 > 159$ mV, Tafel slope > 108 mV dec$^{-1}$, and $i_0 < 1.8 \times 10^{-4}$ mA cm$^{-2}$) of the five data sets generated from the last batch of the ML process (Table S1), validating that the ML model successfully predicted the most optimal conditions and M@CQD catalyst for HER from unknown space.

To gain more insight into the origin of the efficient HER activity of the M@CQD, the electrochemical impedance spectroscopy (EIS) measurements of the electrocatalysts were performed. Figure 6d shows the Nyquist plots with a fitted Randles circuit of electrodes. The charge transfer resistance (Rct) of 189 Ω of Ni@CQD is the lowest among these electrocatalysts, implying that Ni efficiently optimizes the intrinsic catalytic activity. We further evaluate the electrochemical active surface area (ECSA) by extracting the electrochemical double-layer capacitance ($C_{dl}$) from the cyclic voltammetry (CV) curves in the potential range of 1.15–1.25 V (vs. RHE) (Figure 6e). The measured $C_{dl}$ of Ni@CQD is 34.26 mF cm$^{-2}$ which is comparable to that of the other M@CQDs but significantly higher than that of the pristine CQD (7.6 mF cm$^{-2}$) and N-CQD (10.3 mF cm$^{-2}$) (Figure S14). Furthermore, Ni@CQD exhibited the highest turnover frequency (TOF) at the potential of 0.3 V vs. RHE (Figure S14c), which is significantly higher than that of CQD (2.3 s$^{-1}$), Co@CQD (12.7 s$^{-1}$), and Mn@CQD catalysts (11.3 s$^{-1}$).

Ni@CQD also exhibited the lowest ECSA normalized overpotential $\eta_1$ (4.21 mV vs. RHE), compared to $\eta_1$ of CQD (33.2 mV vs. RHE), Co@CQD (4.67 mV vs. RHE), and Mn@CQD (4.89 mV vs. RHE) catalysts.

All measured results demonstrated that TM doping in CQDs improved the electron transport capacity, accelerated HER reaction kinetics on the surface, increased the number of accessible active sites of the catalyst, and enhanced the electrocatalytic performance in the following order: CQD < Mn@CQD < Co@CQD < Ni@CQD.

Electrochemical stability is an essential catalytic property to be evaluated for potential industrial applications. The accelerated degradation measurements of the as-prepared catalysts were performed through continuous CV sweeps at 10 mV s$^{-1}$. The post-potential-cycling LSV curves after 1,000 cycles are obtained, as shown in Figure 6f. The polarization curves before and after 1,000 CV cycles revealed an insignificant change, indicating the excellent durability of the Ni@CQD catalyst. In addition, with a continuous HER at a static overpotential, the current density was maintained well for > 10h (Figure 4h). Table S2 compares the catalysts in this study with previous reports. Note that although such a comparison is important to evalute the efficiency of the in-progressing catalyst, it may not be critical due to various practical factors such as used substrates, pre-calibration before measurement, and impurity of chemicals.[32-34]

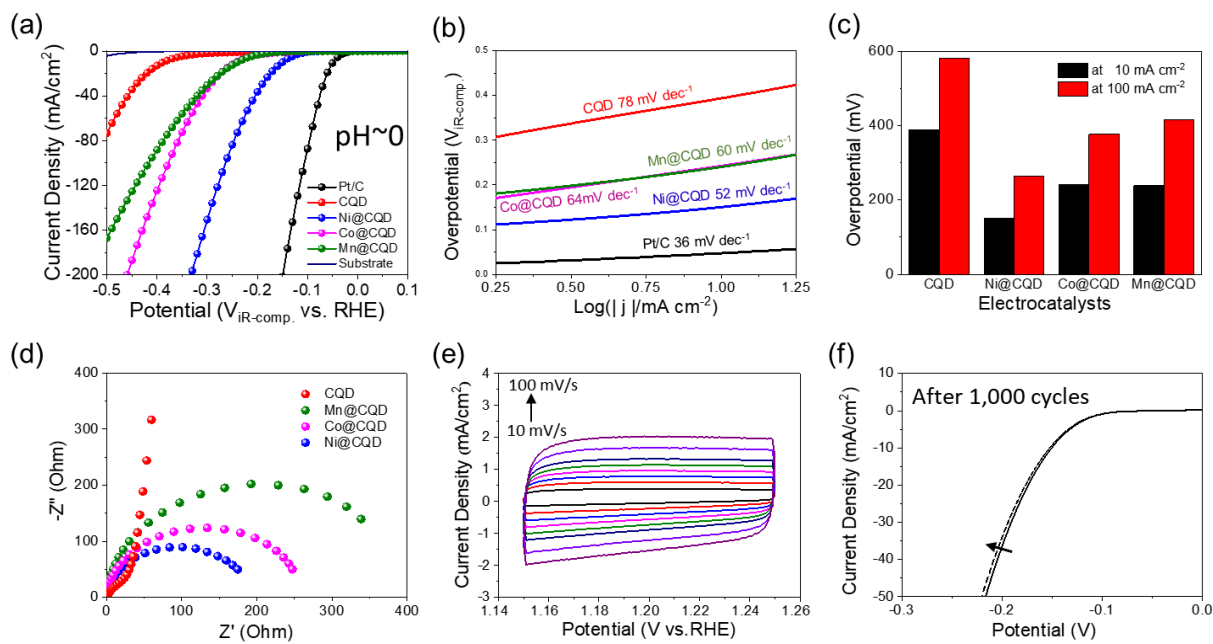

**Figure 6.** Electrochemical analysis in 0.5 M $H_2SO_4$ electrolyte. (a) HER polarization curves of various M@CQD samples. (b) Tafel plots derived from the corresponding HER polarization curves. (c) HER comparison of overpotentials required to achieve current density densities of 10 and 100 mA $cm^{-2}$ vs. RHE for various M@CQD samples. (d) EIS Nyquist plots of catalysts. (e) Electrochemical double layer capacitance extracted from CV curves at various scan rates. (f) Polarization curves for Ni@CQD before and after 1,000 cycles.

Generally, the HER mechanism under acidic conditions is adopted by the hydrogen binding energy (HBE) theory[48] and interface water[49] and/or ion transfer theory.[50] Considering the sluggish HER kinetics following the Volmer-Heyrovsky mechanism in all catalysts, we focused on the contribution of HBE and electrochemical active sites to the HER process. We correlated TM doping with either single atoms or metal clusters on the carbon network based on the observed TMs in the TEM images. The electrochemical results partially demonstrate that the HBE of M@CQD is critical for the catalytic activity on the metal sites in the acidic HER. While the nitrogen and oxygen in the pristine CQDs are the primary proton binding sites, TM with a high oxidation state is energetically favorable for proton adsorption and hydrogen evolution.[51] Such an intrinsic nature of the exposed *d*-block elements would reduce the energy barrier of proton adsorption,[52] which agrees well with the Tafel slope analysis.

**2.3. Density functional theory calculation**

Based on the material characterization and electrochemical measurement results, theoretical investigations were conducted to disclose the effect of the coordinate binding of the TM atom(s) on CQDs toward the electronic structure and HER catalytic activity. Here, we performed DFT simulations for the M@CQD model using three TMs, Ni, Co, and Mn which show good electrochemical performance in the HER experiments. Five different sites for doping TM were considered, as shown in Figure S15 which are named by the amount and type of bonding elements (*e.g.,* 2N-1C associates with the bond between the TM and CQD via two nitrogen and one carbon). Based on the ML and experimental analysis, we constructed models that consist of M@CQD adsorbed to rGO as a conductor and analyzed their interactions with free radical H (Figure 7a). To verify the ML result where 3D rGO was predicted to be more beneficial for the HER than 2D rGO, we constructed two models of rGO conductor as 2D and 3D configurations of carbon network (Figure S16). The calculated parameters of the formation energy and the Gibbs free energy of hydrogen adsorption ($\Delta G_{H^*}$) are represented in Tables S3-4. From the $\Delta G_{H^*}$ results and the formation probability of M@CQD shown in Figure S17, the 2N-1C model using the 3D rGO conductor will be facilely generated with the highest stability. In addition, the weak binding strength and low formation probability of the 2D rGO conductors with the M@CQD catalyst indicate that the M@CQD catalyst is unlikely to bind with 2D rGO, which explains why the ML model predicted the 3D rGO as the optimal conductor. Therefore, we used the 2N-1C M@CQD/3D rGO model for the remaining DFT simulations. The density of states (DOS) of the M@CQD catalyst is shown in Figure 7b. Following the order of Mn < Co < Ni, the height of the DOS peak at the Fermi level and the electrical conductivity present a typical increasing trend. Such enhancement in conductivity results in a higher carrier transfer rate, which is consistent with the previous electrochemically experimental data.

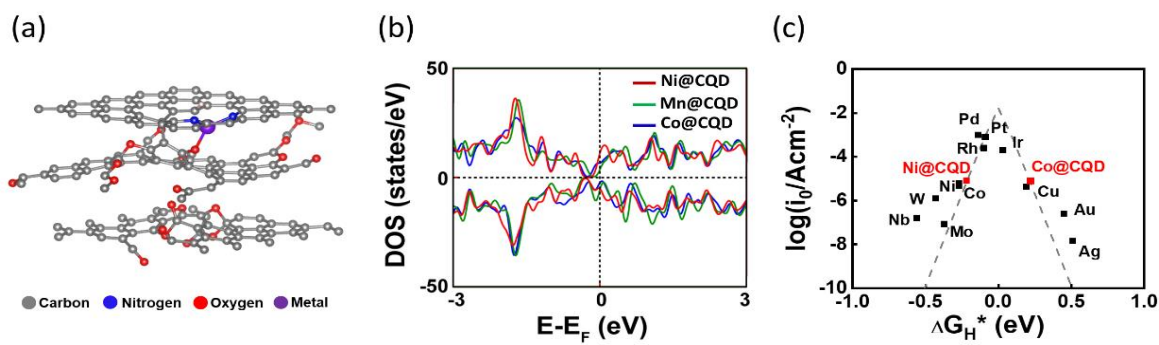

**Figure 7.** DFT simulations of the M@CQD catalysts for HER. (a) Optimized structure of M@CQD interacting with rGO conductor. (b) DOS of M@CQD. (c) The volcano curve of the exchange current density ($i_0$) as a function of the proton adsorption Gibbs free energy.

As the metal d orbital–adsorbate orbital interaction is the most dominant factor for electrocatalytic activity,[52] we examined the hydrogen evolution capability of M@CQD/3D rGO by using the TM in the carbon network as the hydrogen adsorption site. The metal-free CQD has the most inert surface, and the corresponding active site will be associated with the nitrogen/oxygen-containing functional groups at the edge, irrespective of whether there is nitrogen or oxygen in the basal plane. Because Ni exhibits a higher electronegativity than Co and Mn, it is reasonable to assume that the Ni atom in Ni@CQD exhibits the best adsorption of positively charged $H^+$ in an acidic medium among the three TMs. Hence, we calculated the integrated-crystal orbital Hamilton population (ICOHP) of the H-M (M = Ni, Co, Mn) bond for M@CQD/3D rGO. Generally, the more negative the ICOHP value, the stronger the H-M bonding interaction. Regarding, the H-Ni bond has a more negative ICOHP value (−6.787 eV) than the H-Co (−6.403 eV) and H-Mn (−5.981 eV) bonds, demonstrating that the Ni atom will capture the free proton more effectively. Next, a volcano curve is plotted with the theoretical exchange current density ($i_0$) as a function of $\Delta G_{H^*}$ in Figure 5c. For an ideal HER catalyst, $\Delta G_{H^*}$ should be close to zero, in accordance with the Sabatier principle. Although Ni@ and Co@CQD share the same absolute $\Delta G_{H^*}$ value (0.22 eV), Ni@CQD with a negative $\Delta G_{H^*}$ is on the left branch, which is conducive to strong hydrogen adsorption, whereas the Co@CQD is on the right branch, which indicates the hydrogen adsorption is significantly hindered. In the case of Mn@CQD, a highly positive value $\Delta G_{H^*}$ of 0.61 eV was achieved which indicates strong hydrogen desorption of $H_2$ molecules and a lack of ability to adsorb hydrogen. Therefore, the Ni@CQD and Co@CQD catalysts have great potential as superior HER catalysts, whereas the Ni@CQD catalyst is more beneficial in terms of the electronic structure.

## 3. CONCLUSION

In conclusion, we presented an effective strategy of rational design and optimization process for transient metal-incorporated carbon quantum dot (M@CQD) based HER electrocatalysts that can be readily used in an average electrochemistry laboratory. We also demonstrated that the Ni@CQD catalyst exhibits the best performance under optimal experimental conditions. Our methodology combines a BGA prediction model with verification steps of electrochemical experiments and theoretical DFT simulations to determine the most favorable TM dopant and optimal experiment conditions toward an efficient HER electrocatalyst. Among six earth-abundant TMs, Ni renders the highest performance in exhibiting corresponding electrochemical properties, and the optimal experimental conditions were predicted to be a catalyst loading of 0.8 mg, a 3D rGO conductor, a Ni foam electrode, a temperature of 65°C, and pH 0. The Ni@CQD catalyst exhibited the lowest onset potential of 135 mV and overpotential of 189 mV at 10 mA cm$^{-2}$, the lowest Tafel slope of 52 mV dec$^{-1}$, and the lowest Rct of 189 Ω, in additon to excellent cyclability and long-term electrocatalytic stability. DFT simulation results rationalized why the 3D rGO is the most optimal conductor for the HER as the Ni@CQD/3D rGO hybrid structure exhibited better stability than the Ni@CQD/2D rGO. In addition, the Ni@CQD catalyst exhibited the best HER properties in terms of electronic structure, verifying the ML predictions. We expect that the excellent performance of Ni@CQD provides an effective route for the rational design of highly active HER electrocatalysts for commercial applications, as well as our proposed method can significantly accelerate the development of efficient electrocatalysts not only for the HER, but also for other electrochemical reactions.

## 4. MATERIALS AND METHODS

**4.1. Synthesis of single Transition Metal doped Carbon Quantum Dots**

CQDs were synthesized via a facile solvothermal reaction, as mentioned in previous reports.[53] Briefly, 300 mg of fumaronitrile was dispersed in 3.0 mL of de-ionized water. The hydrothermal reaction was conducted at 200°C for 20 min. The product, as the CQD, was obtained via repeated filtering with a polytetrafluoroethylene (PTFE) membrane and centrifugation/re-dispersion in water. The same reaction was used to synthesize r-CQD, but tetrahydrofuran (THF) was used instead of de-ionized water. To synthesize M@CQD, a small amount of metal precursor (*i.e.*, metal acetate or metal nitrate) was added simultaneously during the reaction with a molar ratio to fumaronitrile of 3%. The hydrothermal reaction was performed under typical conditions, and the obtained CQDs were stored in a refrigerator for further use.

**4.2. Material characterization**

Wide-angle powder XRD patterns were measured using a Bruker D8 Advance diffractometer. The morphologies, microstructure and elemental components of the hybrids were investigated by field emission scanning electron microscopy (FESEM, JSM-7600F, JEOL), and HRTEM (Cs-corrected HRTEM, JEM ARM 200F) equipped with energy dispersive spectrometry (EDS) at an acceleration voltage of 200 kV, respectively. XPS was performed using a Thermo ESCALab 250 X-ray photoelectron spectrometer.

**4.3. Electrochemical measurements**

All electrochemical measurements were performed using an electrochemical workstation CHI 660B in a standard three-electrode set-up system in a 0.5 M $H_2SO_4$ aqueous solution. To prepare the catalyst ink, a 4 mg catalyst, 0.5 mg conductor 3D rGO and 15 μL of Nafion (5 wt%, Alfa Aesar) were dispersed in 1 mL of water/ethanol mixture solution (1:1 v:v), followed by ultrasonication for 30 min. Then, 7.5 μL of the catalyst ink was dropped on a nickel foam electrode. The prepared sample was then dried in a vacuum oven under 15 MPa of pressure overnight. The benchmark Pt/C (20 wt%) was used as a reference catalyst for the HER test. Before the HER test, the electrochemical system was calibrated by using Pt wire as the working and reference electrodes. LSV was performed at a 10 mV $s^{-1}$ scan rate using an Ag/AgCl reference electrode and a graphite rod counter electrode. All LSV polarization curves were corrected by iR compensation. EIS measurements were conducted in the frequency range of 100 kHz-0.1 Hz with an overpotential of 140 mV. The durability measurements were performed by CV sweep 1,000 times with a scan rate of 10 mV $s^{-1}$. All reported potentials were in accordance with the RHE.

TOF values were calculated based on the weight ratio of CQDs to TM as:

$$TOF = I / (2Fn),$$

where, $I$ denotes the current, $F$ denotes the Faraday constant (96 485 C $mol^{-1}$), and $n$ denotes the number of active sites (mol). We assume that two electrons are required to form one hydrogen molecule from two protons, thus the denominator is multiplied by 2.

For the number of active sites $n$, we assume that the active site of pristine CQD is generated from nitrogen atoms at edge-sites while the active site of M@CQD could be considered as the TMs doped in CQD. Thus, the number of active sites equals the number of M atoms in M@CQD, which is estimated from the XPS elemental analysis (Figure 4e).

### 4.4. Theoretical calculations

DFT calculations were performed using the Vienna ab initio Simulation Package (VASP) code.[54] The projector-augmented-wave pseudopotential was employed for all calculations to treat the core electrons,[55] with a cut-off energy of 400 eV. The Perdew-Burke-Ernzerhof (PBE) approach of spin-polarized generalized gradient approximation (GGA) functional was used to describe the exchange-correlation energy.[56] The convergence tolerance of energy and force was $10^{-5}$ eV and 0.01 eV/Å, respectively. The van der Waals interactions were applied using the empirical correction in Grimme's scheme (DFT-D3).[57]

The formation energy of M@CQD was defined by:

$$E_f = E_{M@CQD} - E_{CQD} - E_M,$$

where $E_{M@CQD}$, $E_{CQD}$, and $E_M$ denote the total energy of M@CQD, CQD, and the TM dopant, respectively.

The hydrogen adsorption free energy of a catalyst was defined by:

$$\Delta G_{H^*} = \Delta E + \Delta ZPE + T\Delta S,$$

where $E$ denotes the electronic energy, $ZPE$ denotes the zero-point energy, $T$ denotes the temperature, and $S$ denotes the entropy. Δ denotes the difference between the H-adsorbed surface and a pristine surface.

**4.5. Machine learning-based motif**

We attempted two approaches to reduce the number of experiments required for optimization. First, we categorized all independent variables. Second, the Bayesian approach was applied to the GA for catergorical independant variables. In other words, the posterior probability was used when performing the GA crossover and mutation operations. In this study, the uniform crossover was used, and when selecting a new value of each variable for the new generation, it was sampled in a distribution proportional to the posterior probability. In the mutation operation, when the index of a variable of the previous generation is i, the index of the next generation variable was sampled to be proportional to the posterior probability of the variable among i+1/i/i-1. After obtaining the cost function of the new batch from experiments, the new posterior probability was calculated using the exponential weight average from the posterior probability up to the previous batch.

ACKNOWLEDGMENTS

This work was supported by the National Research Foundation of Korea (NRF) grant funded by the Korean government (MSIT) (NRF-2022R1A2C1011559).

Received: ((will be filled in by the editorial staff))
Revised: ((will be filled in by the editorial staff))
Published online: ((will be filled in by the editorial staff))

CONFLICT OF INTEREST

The authors declare no conflict of interest.

ORCID

Duong Nguyen Nguyen - http://orcid.org/0000-0002-7455-2241

Min-Cheol Kim - https://orcid.org/0000-0002-4457-7421

Ho Seok Park - https://orcid.org/0000-0002-4424-4037

Uk Sim - https://orcid.org/0000-0001-7767-495X

Jung Kyu Kim - http://orcid.org/0000-0002-8218-0062

A novel data-driven strategy for the design and optimization of earth-abundant transition metal-doped carbon quantum dot toward hydrogen evolution reaction is proposed. Using machine learning for catalyst prediction and optimization, experimental and theoretical verification are performed to verify the accuracy of this approach. Consequently, a full picture of HER performance of M@CQD within the parameter space has been provided.

**Machine learning, carbon quantum dot, transition metal doping, hydrogen evolution reaction, density functional theory**

Duong Nguyen Nguyen,[1] Min-Cheol Kim,[1] Unbeom Baeck,[1] Jaehyoung Lim,[2] Namsoo Shin,[3] Jaekook Kim,[4] Heechae Choi,[5] Ho Seok Park,[1,6,*], Uk Sim,[2,7,*] and Jung Kyu Kim[1,*]

**Machine Learning Assisted Design and Optimization of Transition Metal-Incorporated Carbon Quantum Dot Catalysts for Hydrogen Evolution Reaction**

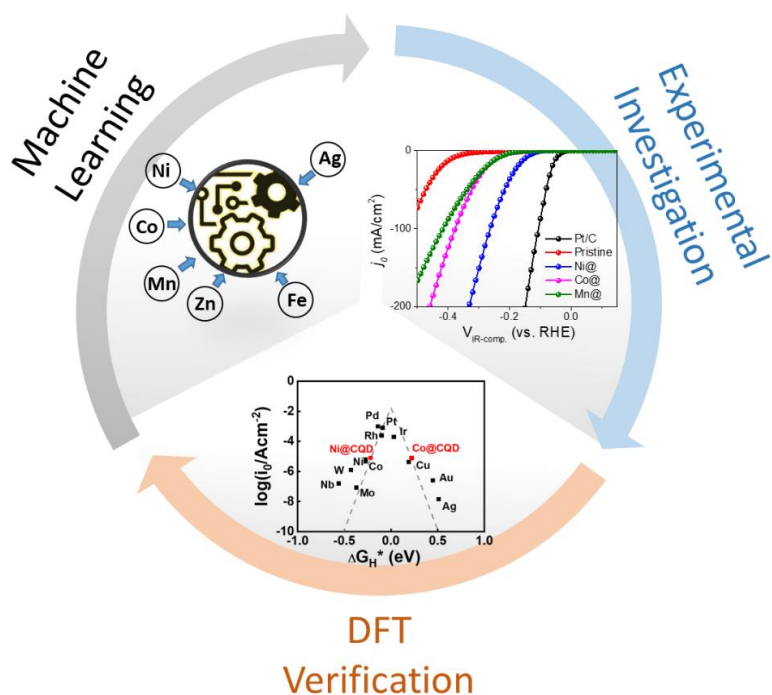

# Supporting Information

**Machine Learning Assisted Design and Optimization of Transition Metal-Incorporated Carbon Quantum Dot Catalysts for Hydrogen Evolution Reaction**

Duong Nguyen Nguyen,[1] Min-Cheol Kim,[1] Unbeom Baeck,[1] Jaehyoung Lim,[2] Namsoo Shin,[3] Jaekook Kim,[4] Heechae Choi,[5] Ho Seok Park,[1,6,*], Uk Sim,[2,7,*] and Jung Kyu Kim[1,*]

[1]School of Chemical Engineering, Sungkyunkwan University (SKKU), 2066, Seoburo, Jangan-gu, Suwon 16419, Republic of Korea
[2]Hydrogen Energy Technology Laboratory, Korea Institute of Energy Technology (KENTECH), Naju 58330, Republic of Korea
[3]Deep Solution Inc., 17, Jukjeon ro, Giheunggu, Yonginsi, Gyeonggido 16897, Republic of Korea
[4]Department of Materials Science and Engineering, Chonnam National University, Gwangju, 61186, Republic of Korea
[5]Theoretical Materials & Chemistry Group, Institute of Inorganic Chemistry, University of Cologne, Greinstr. 6, 50939 Cologne, Germany
[6]SKKU Institute of Energy Science and Technology (SIEST), Sungkyunkwan University (SKKU), Suwon, 16419, Republic of Korea
[7]Research Institute, NEEL Sciences, INC., Gwangju, 61186, Republic of Korea

**Correspondence**
Jung Kyu Kim, School of Chemical Engineering, Sungkyunkwan University (SKKU), 2066, Seoburo, Jangan-gu, Suwon 16419, Republic of Korea
Email: legkim@skku.edu

Uk Sim, Hydrogen Energy Technology Laboratory, Korea Institute of Energy Technology (KENTECH), Naju, Republic of Korea. Research Institute, NEEL Sciences, Inc., Naju, Republic of Korea
Email: usim@kentech.ac.kr

Ho Seok Park, School of Chemical Engineering, SKKU Institute of Energy Science and Technology (SIEST), Sungkyunkwan University (SKKU), Suwon, 16419, Republic of Korea
Email: phs0727@skku.edu

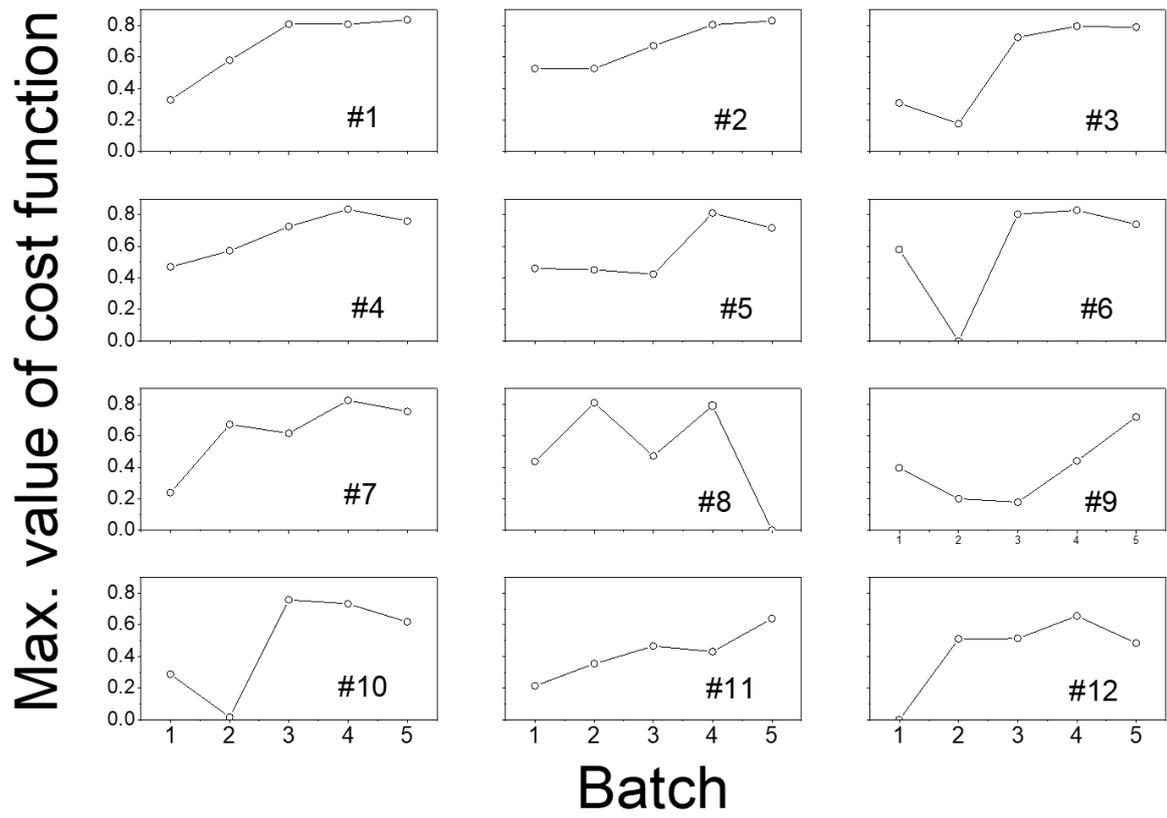

**Figure S1.** Cost function values using bayesian genetic algorithm of all batch data.

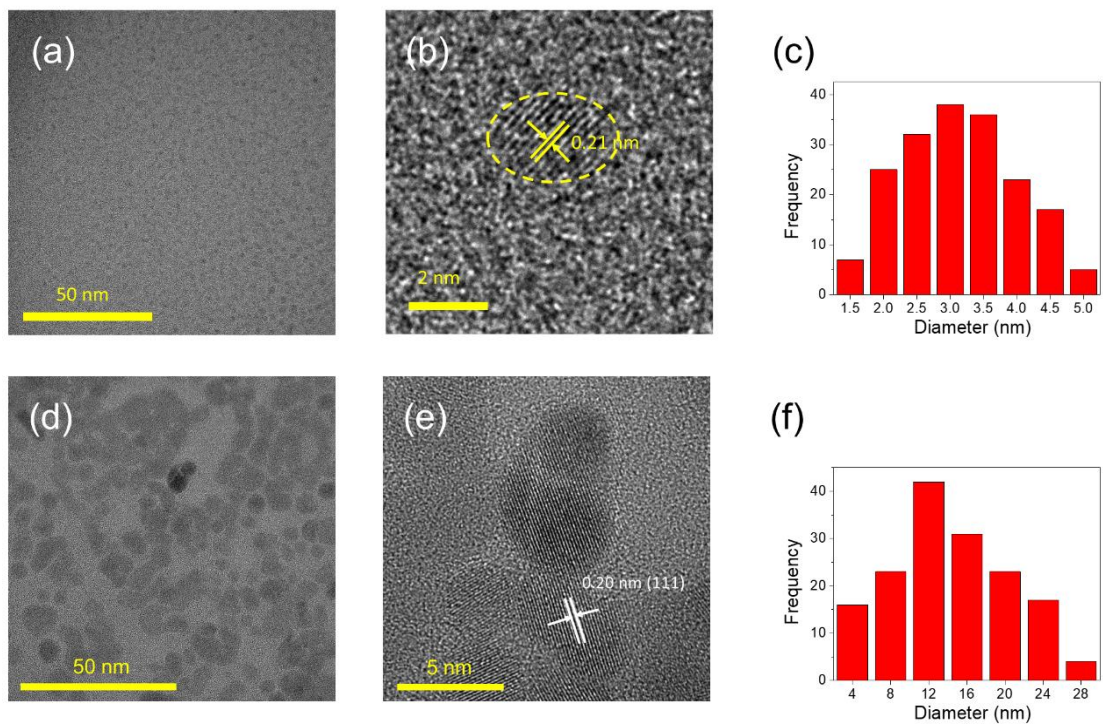

**Figure S2.** TEM images and size distribution of obtained products synthesized by the protocol in this work with the molar ratio of transition metal dopant to carbon precursor as (a) 0%, and (b) 20%.

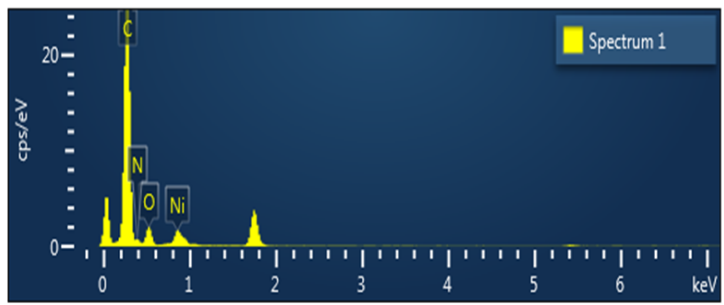

| Element | Wt% | Atomic % |
|---|---:|---:|
| C | 91.70 | 94.96 |
| N | 1.01 | 0.89 |
| O | 4.60 | 3.57 |
| Ni | 2.70 | 0.57 |
| Total: | 100.00 | 100.00 |

**Figure S3.** Energy-dispersive X-ray (EDX) spectroscopy results and percentage amount of typical elements in Ni@CQD sample.

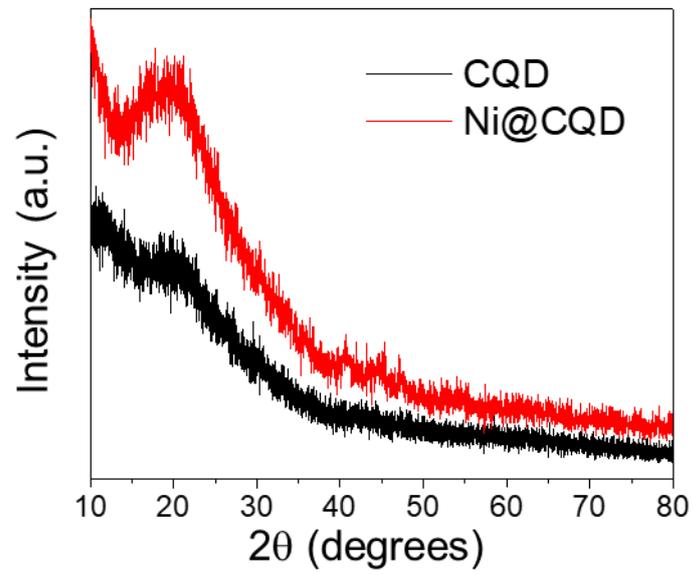

**Figure S4.** XRD pattern of pristine CQD and Ni@CQD.

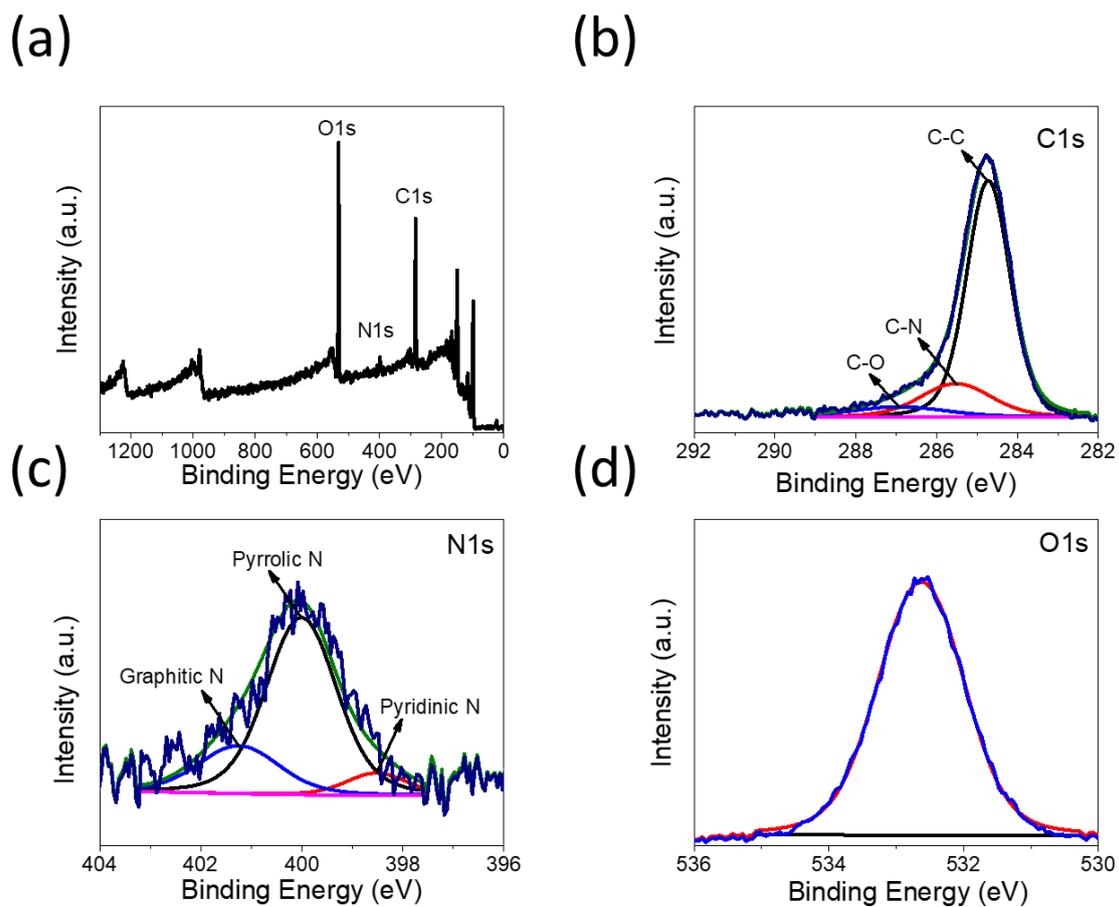

**Figure S5.** (a) XPS survey and corresponding high resolution XPS spectra of (b) C 1s, (c) N 1s, and (d) O 1s of pristine CQD sample.

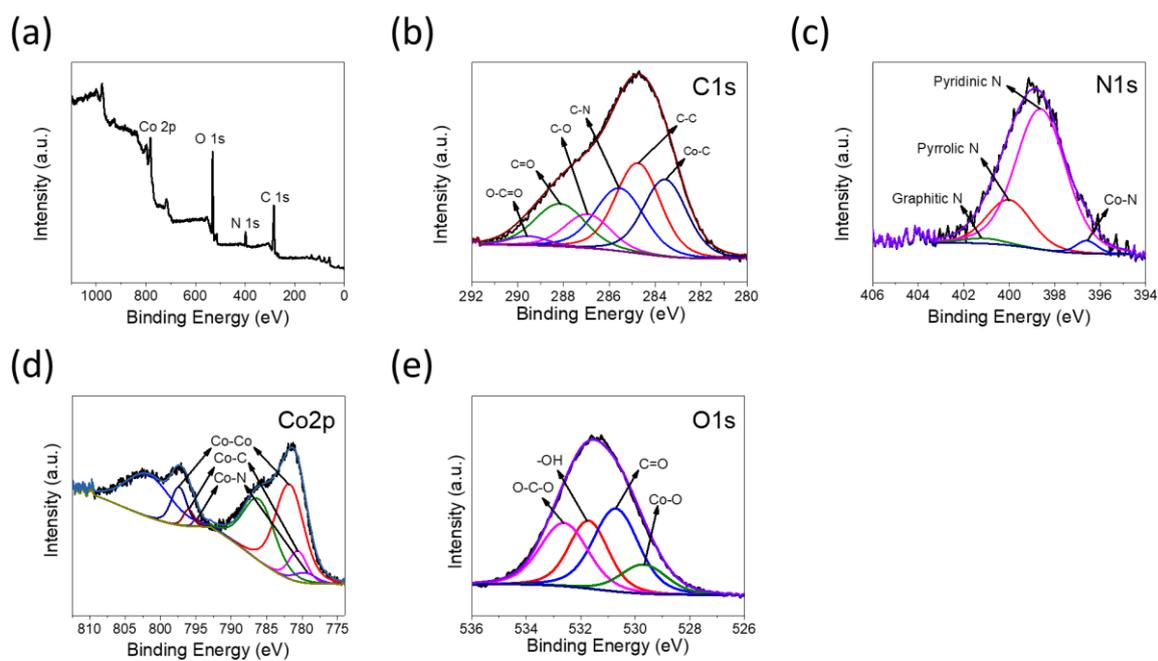

**Figure S6.** (a) XPS survey and corresponding high resolution XPS spectra of (b) C 1s, (c) N 1s, (d) Co 2p and (e) O 1s of Co@CQD sample.

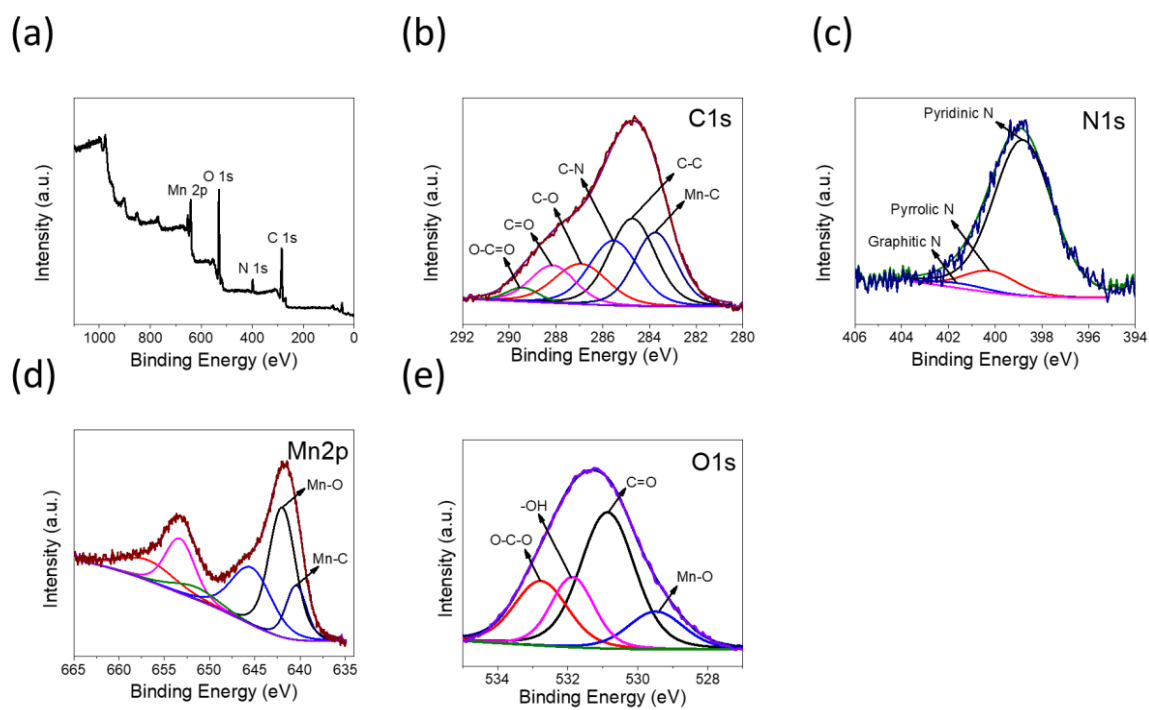

**Figure S7.** (a) XPS survey and corresponding high resolution XPS spectra of (b) C 1s, (c) N 1s, (d) Mn 2p and (e) O 1s of Mn@CQD sample.

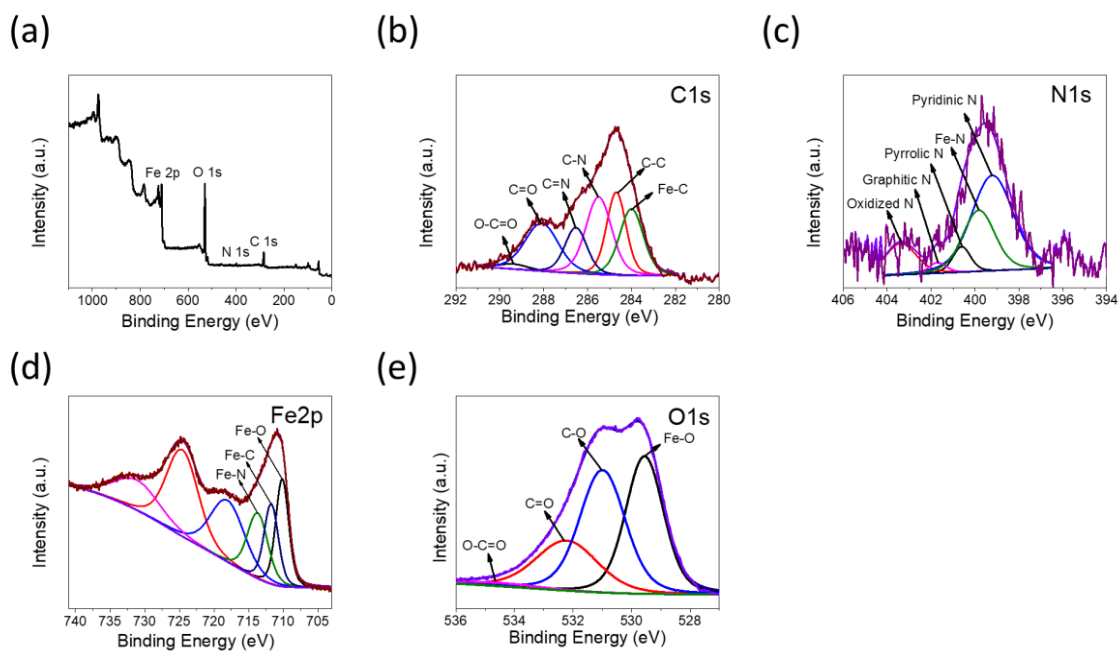

**Figure S8.** (a) XPS survey and corresponding high resolution XPS spectra of (b) C 1s, (c) N 1s, (d) Fe 2p and (e) O 1s of Fe@CQD sample.

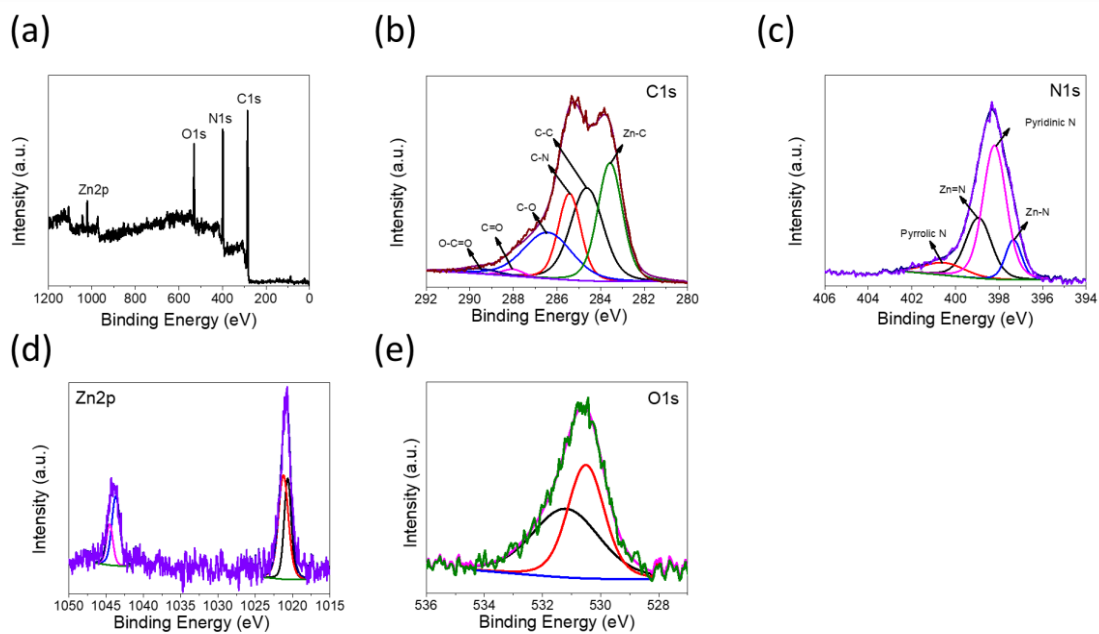

**Figure S9.** (a) XPS survey and corresponding high resolution XPS spectra of (b) C 1s, (c) N 1s, (d) Zn 2p and (e) O 1s of Zn@CQD sample.

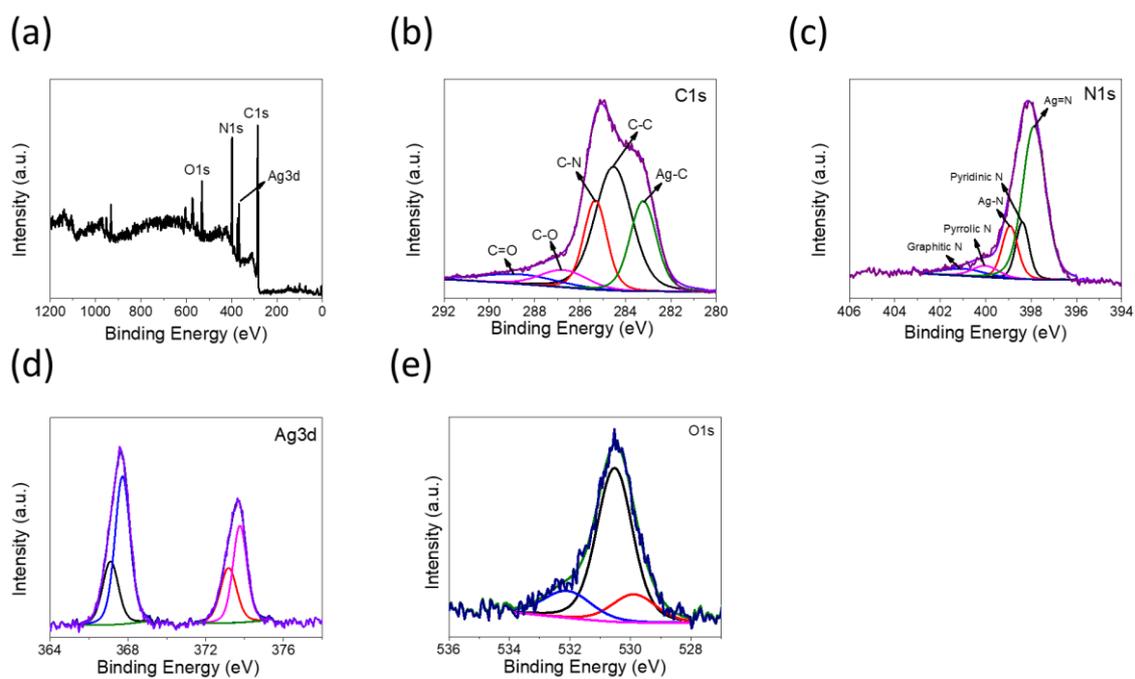

**Figure S10.** (a) XPS survey and corresponding high resolution XPS spectra of (b) C 1s, (c) N 1s, (d) Ag 3d and (e) O 1s of Ag@CQD sample.

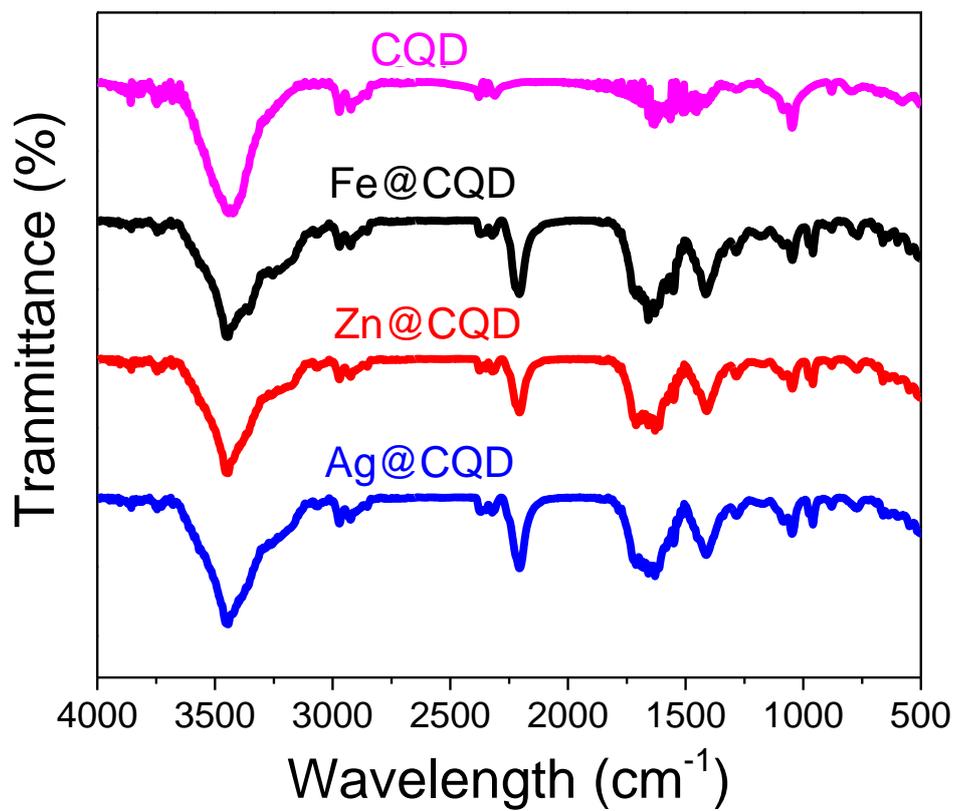

**Figure S11.** FT-IR spectra of as-prepared CQDs.

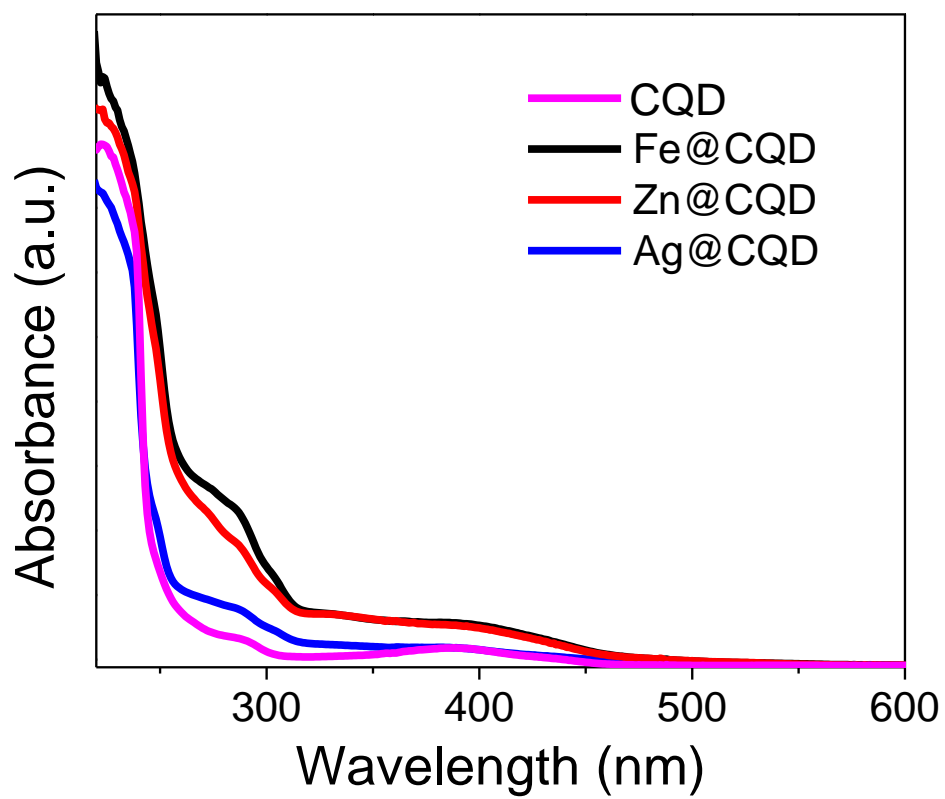

**Figure S12.** UV-Vis spectra of as-prepared CQDs.

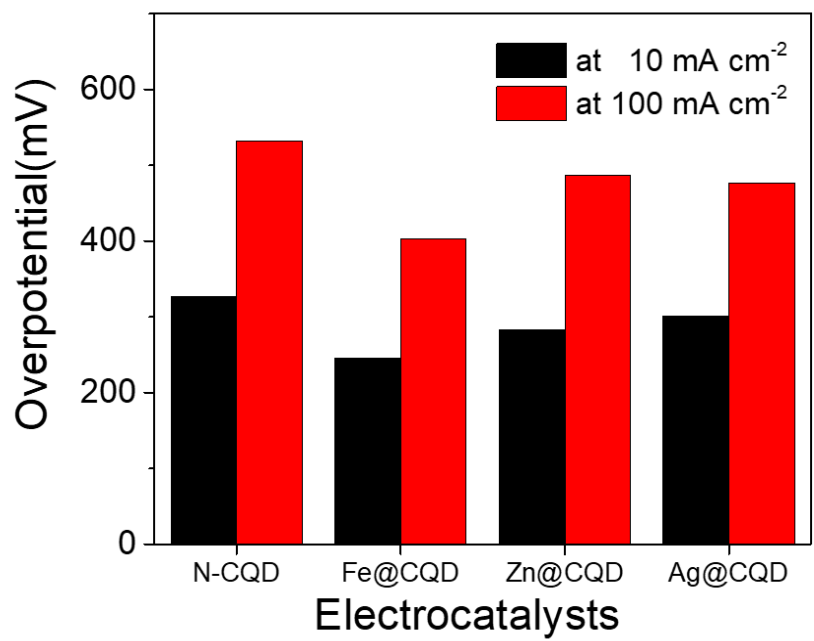

**Figure S13.** HER comparison of overpotentials required to achieve a current densities of 10 and 100 mA/cm$^2$ vs. RHE for various M@CQD samples.

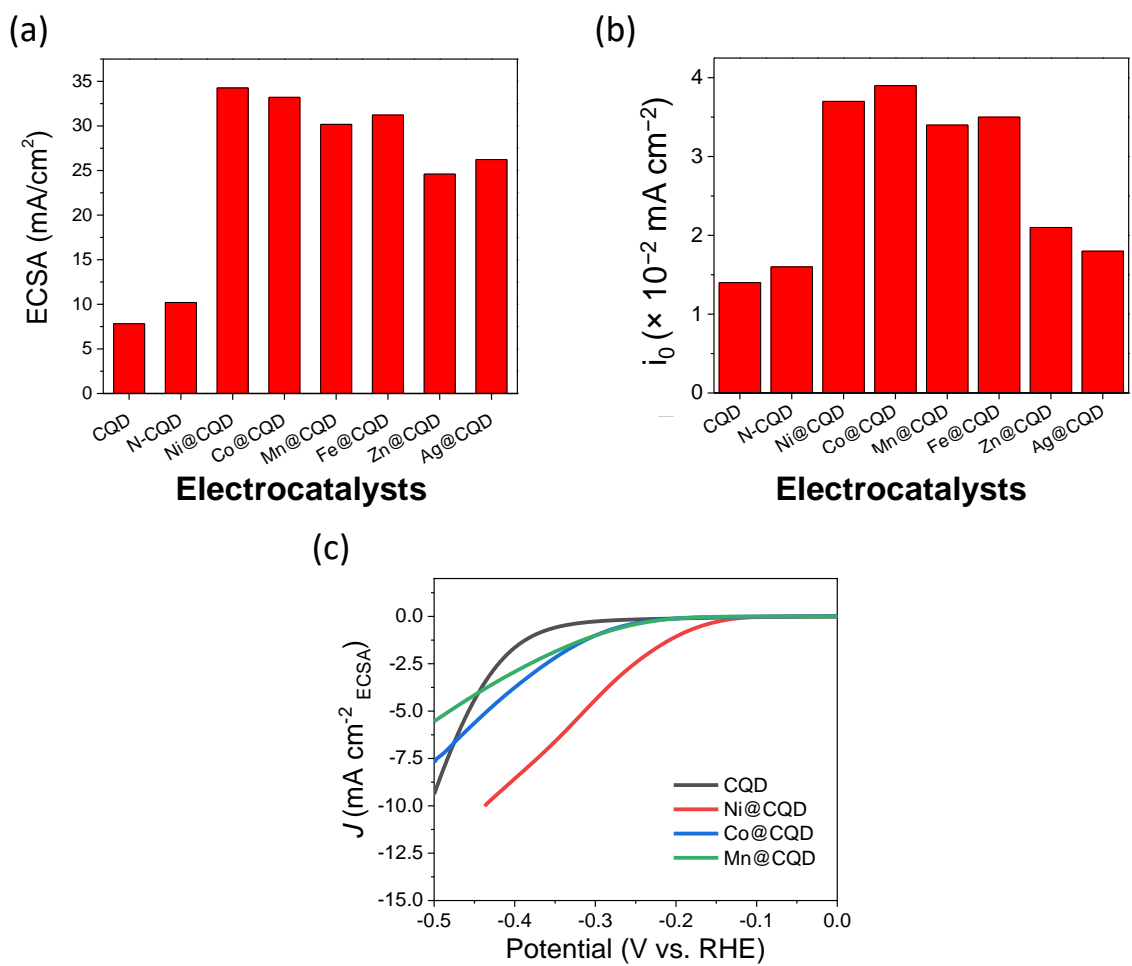

**Figure S14.** HER comparison of (a) ECSA and (b) exchange current density for various M@CQD samples. (c) HER Polarization curve normalized to the ECSA for various M@CQD samples.

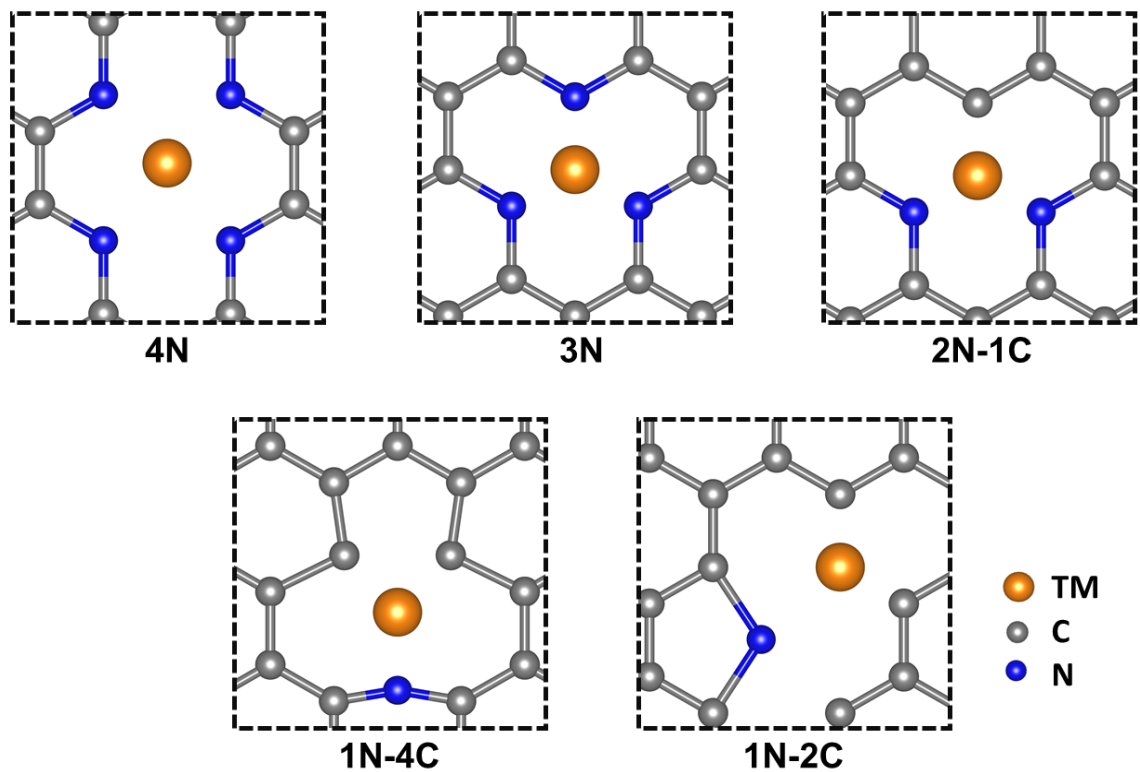

**Figure S15.** Five proposed models for bonding between transition metal (TM) and the carbon network via carbon (C) and nitrogen (N) atoms.

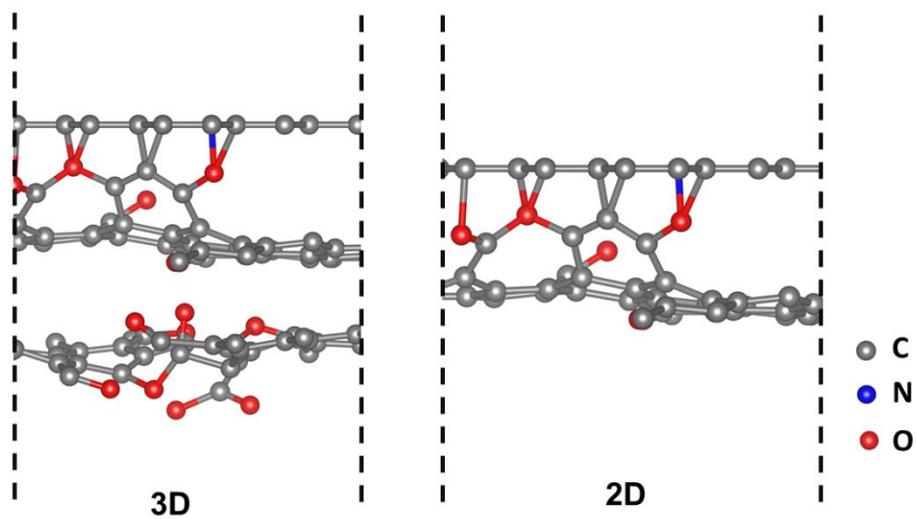

**Figure S16.** Two proposed models for CQD simulation as two dimensional (2D) and three dimensional (3D) carbon networks.

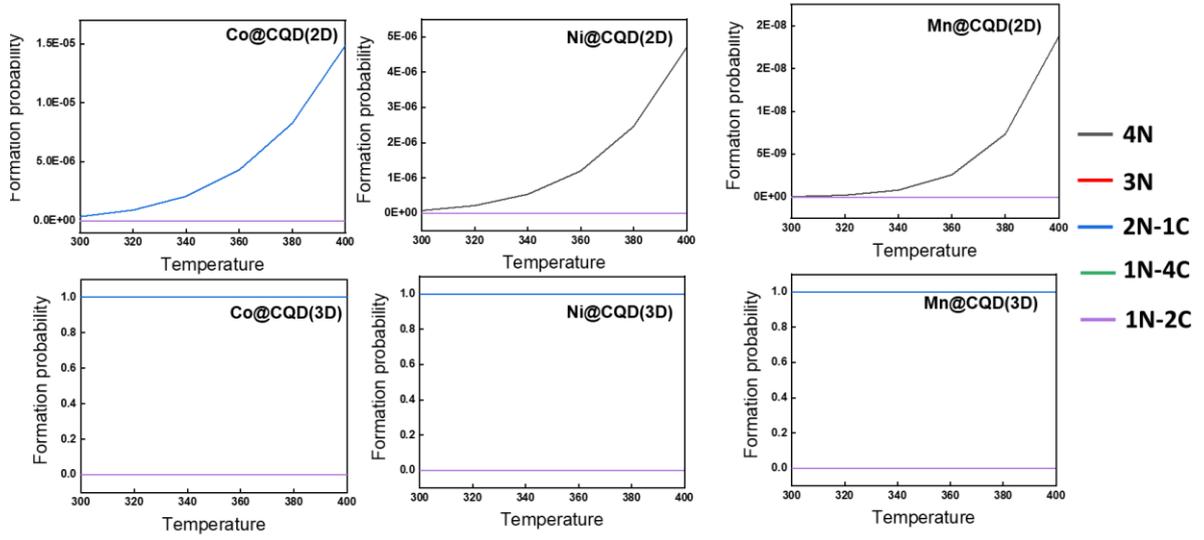

**Figure S17.** Formation probability of M@CQD on rGO conductor using equation as $P_j = \frac{\exp(-\frac{E_f}{k_b T})}{1+\sum_i^n \exp(-\frac{E_f}{k_b T})}$ with $E_f$, $k_b$, and T denoting the formation energy, Boltzmann constant and temperature, respectively.

**Table S1.** Electrochemical measurements of the five data sets from the last batch of the ML model compared with those of Ni@CQD.

|  | **Overpotential $\eta_1$ (mV@1 mA cm$^{-2}$)** | **Tafel slope (mV dec$^{-1}$)** | **Exchange current density $i_0$ (mA cm$^{-2}$)** |
| --- | --- | --- | --- |
| **Ni@CQD** | **135** | **52** | **3.7 × 10$^{-2}$** |
| Data set #1 | 159 | 108 | 1.1 × 10$^{-4}$ |
| Data set #2 | 195 | 111 | 0.8 × 10$^{-4}$ |
| Data set #3 | 220 | 117 | 1.3 × 10$^{-4}$ |
| Data set #4 | 267 | 142 | 1.7 × 10$^{-4}$ |
| Data set #5 | 218 | 164 | 1.8 × 10$^{-4}$ |

**Table S2.** HER performance of CQD-based electrocatalysts reported in previous literature.

| Catalyst | Electrode | Electrolyte | Tafel slope (mV/dec) | $\eta_{10}$ (mV) | Ref. |
|---|---|---|---|---|---|
| N-CQD | Graphite foam | 0.5M $H_2SO_4$ | 104 | 237 | *Chem. Eng. Sci., 194 (2019), pp. 54-57* |
| CQD@NiCoP | Glass carbon | 0.5M $H_2SO_4$ | 80 | 108 | *Chemical Engineering Journal* 351 (2018): 189-194 |
| CQD@NiMoP | Glass carbon | 0.5M $H_2SO_4$ | 41 | 183 | Industrial & Engineering Chemistry Research, 58(31), 14098-14105 |
| CQD/MoP | Glass carbon | 1M KOH | 56 | 210 | ACS applied materials & interfaces, 10(11), 9460-9467 |
| CQD/CoS | Glass carbon | 0.5M $H_2SO_4$ | 56 | 165 | J. Mater. Chem. A, 5 (2017), pp. 2717-2723 |
| N-CQD/$Ni_3S_2$ | Nickel foam | 1M KOH | 95.5 | 218 | Small, 13(24), 1700264 |
| CQD/$MoS_2$ | Glass carbon | 0.5M $H_2SO_4$ | 43 | 200 | Electrochim. Acta, 211 (2016), pp. 603-610 |
| CQD@Au | Glass carbon | 0.5M $H_2SO_4$ | 78 | 130 | Chem. Phys. Lett., 641 (2015), pp. 29-32 |
| $Co_9S_8$@CQD | Glass carbon | 1M KOH | 101.8 | 261 | RSC Adv., 7 (2017), pp. 19181-19188 |
| $NiCo_2P_2$/CQD | Graphite plate | 1M KOH | 62.3 | 119 | Nano Energy, 48 (2018), pp. 284-291 |
| **Ni@CQD** | **Nickel foam** | **0.5M $H_2SO_4$** | **52** | **189** | **This work** |

**Table S3.** Calculated parameter of formation energy of M@CQD with 3D rGO.

|       | $E_f$ (eV) |       |       | $\Delta G_{H^*}$ (eV) |       |       |
|-------|-------|-------|-------|-------|-------|-------|
| 4N    | -1.89 | -1.88 | -2.70 | 0.19  | 1.21  | 0.54  |
| 3N    | -0.91 | -0.66 | -1.17 | 0.01  | -0.20 | -0.18 |
| 2N-1C | -3.06 | -2.40 | -3.35 | 0.22  | -0.22 | 0.61  |
| 1N-4C | -2.43 | -1.82 | -2.47 | 0.03  | 0.06  | -0.17 |
| 1N-2C | 9.61  | 9.72  | 8.63  | -0.03 | -0.61 | 0.54  |

**Table S4.** Calculated parameter of formation energy of M@CQD with 2D rGO.

|       | $E_f$ (eV) |       |       | $\Delta G_{H^*}$ (eV) |       |       |
|-------|-------|-------|-------|-------|-------|-------|
| 4N    | -2.02 | -1.97 | -2.74 | 0.19  | 0.93  | 0.52  |
| 3N    | -0.03 | 0.10  | -0.49 | 0.15  | 0.50  | 0.27  |
| 2N-1C | -2.68 | -1.36 | -3.02 | 0.21  | 0.69  | 0.64  |
| 1N-4C | -1.29 | -1.00 | -1.33 | -0.03 | 0.03  | -0.04 |
| 1N-2C | 12.52 | 9.42  | 8.77  | -0.46 | -0.61 | 0.52  |